\documentclass[%
 preprint,
superscriptaddress,
 amsmath,amssymb,
  aps,
pra,
]{revtex4-2}

\usepackage{graphicx}
\usepackage{dcolumn}
\usepackage[normalem]{ulem}
\usepackage{bm}

\usepackage{physics}
\usepackage{xcolor}
\usepackage{hyperref}

\usepackage{amsmath,amssymb}

\usepackage{siunitx}

\usepackage{amsmath}
\usepackage{amssymb}

\usepackage{soul}

\usepackage{lipsum}
\usepackage{mathtools}

\begin{document}

\title{An effect of a pump pulse rising edge on the QCL build-up time: the analytical approach }

\author{Ivan I. Vrubel}
\email{ivanvrubel@ya.ru}
\affiliation{Ioffe Institute, Politekhnicheskaya ul. 26, 194021 St. Petersburg, Russia}

\author{Evgeniia D. Cherotchenko}
\affiliation{Ioffe Institute,  Politekhnicheskaya ul. 26, 194021 St. Petersburg, Russia}

\author{Ksenia V. Kusakina}
\affiliation{Lobachevsky University, Nizhny Novgorod, Russia}
\affiliation{Lomonosov Moscow State University, Moscow}

\author{Sausan H. Abdulrazak}
\affiliation{Ioffe Institute, Politekhnicheskaya ul. 26, 194021 St. Petersburg, Russia}

\author{Vladislav V. Dudelev}
\affiliation{Ioffe Institute, Politekhnicheskaya ul. 26, 194021 St. Petersburg, Russia}

\author{Grigorii S. Sokolovskii}
\affiliation{Ioffe Institute, Politekhnicheskaya ul. 26, 194021 St. Petersburg, Russia}

\begin{abstract}
In this work, we provide a simplified theoretical analytical estimation of the quantum cascade laser build-up time, accurately taking into account the main effects: the QCL overheating during the pump pulse and the photon mode filling effect. The non-trivial interplay of the mentioned effects brings about a variety of possible experimental build-up time behaviors. The latter range from decreasing curvature for low-power devices to non-monotonous function for devices performing at high power.
\end{abstract}

\maketitle

\section{Introduction}

Quantum cascade lasers (QCLs) operating in the mid-infrared range are important components for security systems \cite{Kumar2011,grasso2016defence, patel2012qcl}, environmental monitoring \cite{jahjah2014compact, zifarelli2022multi, panda2022ec}, biomedical applications \cite{panda2022ec, abramov2019quantum,ghorbani2017real}, and for wireless optical communication \cite{yang2021room, ashok2021performance, pang2021free}. Such a wide range of applications is associated with the unique properties of the mid-IR domain, in which there are intense absorption lines of various molecules\cite{caffey2011recent}, as well as two atmospheric transparency windows\cite{gmachl2001recent}. QCLs operate on multiple electron intraband optical transitions between quantization levels of the conduction band, carried out by one electron tunneling from one quantum cascade to another when an external electric field is applied\cite{kazarinov1971}. In this case, the lasing wavelength in a QCL is determined by the height of the barriers and the width of the quantum well in which the optical transition occurs. Therefore, based on the same materials forming a quantum well/barrier heteropair, it is possible to create devices for the entire mid-infrared range.
However, the important problem of the QCLs active region overheating arises due to high operating currents and voltages.
Experimentally, one observes an increase in the threshold current \cite{dudelev24} and a decrease in the output power with temperature. Apart from that, the thermal-related processes no less affect the QCL dynamics and, first of all, the build-up time. Theoretical estimates have shown that its value for strong-signal modulation \cite{Hamadou09} is in the range of hundreds picoseconds. At the same time, experimental studies show that this time can exceed 5~ns and significantly depends on the pump pulse edge\cite{Cherotchenko22obs}.
In this work we consider different pump pulse rising edges, theoretically predict their influence on the QCL operation conditions, and in several steps we explicitly discuss the origins of nontrivial build-up time, measured in \cite{Cherotchenko22obs}. With a lack of reliable external modulation technique in mid-IR, this study opens up good prospects for data transmission by direct current modulation.

\section{Results}

In the prominent theoretical research \cite{Hamadou09}, the authors document the build-up time for pumping utilizing a Heaviside rising edge. However, in a real case, especially for high-power devices, the pump pulse rise time is of the order of nanoseconds, which cannot be neglected. In this section, we aimed at the assessment of fundamental limitations that appear in realistic modeling of QCL dynamics.

\subsection{Non-heating case and non-zero rise time}

Consider the effects resulting only from the non-zero rise time of a pump pulse. First, we start with the simplest case, when the temperature of an active region remains constant during the pump. In this subsection, we neglect physical processes occurring inside the active region and assume that emission intensity is simply proportional to the difference between the given pump current value and constant threshold current. 

By definition, the maximum light power in the heating-less case can be calculated as:

\begin{equation}
L_0=\chi (I_0-I_{theor})
\end{equation}

\noindent where $\chi$ is a light current characteristic slope, $I_0$ is a steady pump current, $I_{theor}$ is a constant threshold current in non-heating case. This maximum value is a reference for the moment ($\tau_\alpha$) when the time dependent light power reaches ``$\alpha$'' fraction of equilibrium intensity in the following form: 

\begin{equation}
L(\tau_\alpha)=\chi(I_p(\tau_\alpha)-I_{theor})= \alpha L_0 =  \alpha \chi (I_0-I_{theor})
\end{equation}

\noindent where $I_p(t)$ is the time dependent pumping. 
To assess $\tau_\alpha$, we simplify and rewrite the latter expression and get an equation determining the moment when the difference between pumping and threshold currents reaches the value allowing emission with the $\alpha$ fraction of equilibrium intensity.

\begin{equation}
I_p(\tau_{\alpha})-I_{theor} =\alpha*I_{theor}(n-1)
\label{eq:eqtaualpha}
\end{equation}

\noindent where n=$I_0$/$I_{theor}$ is a ratio of steady pump current to constant threshold current ($n \ge 1$). 

To proceed, we need to introduce the realistic pump current time-dependent profile. Without loss of generality, we adopt the following curve:

\begin{equation}
I_p(t)=I_0(1-e^{-\frac{t}{RC}})
\label{eq:realisticpump}
\end{equation}

\noindent where $RC$ is a time constant of pump source. .

Taking into account the pulse time dependence, one can easily find the moment when current reaches a value able to produce the ``$\alpha$'' fraction of maximum light power:

\begin{equation}
n I_{theor} (1-e^{-\frac{\tau_{\alpha}}{RC}}) - I_{theor}=\alpha I_{theor}(n-1)
\end{equation}

\noindent which after simplification reads:

\begin{equation}
\tau_{\alpha}=-RC \cdot ln(\frac{n-1}{n}) - RC \cdot ln(1-\alpha).    
\end{equation}

The presented formula comprising the logarithm is natural for devices with a static threshold pumped by a power source characterized by exponential growth with a characteristic time constant. 

Assuming that $\alpha$ is small, we simplify the expression further and come to the simple dependence:

\begin{equation}
\tau_{\alpha}=-RC \cdot ln(\frac{n-1}{n}) + \alpha RC
\end{equation}

The build-up time $t_\alpha$ is conventionally a difference between the moment when intensity reaches some percentage of its maximum value ($\alpha$ in our case) and the moment when current reaches its threshold value ($\tau_{theor}$).

The adopted curvature of the pump pulse suggests the $\tau_{theor}$ is a root of the equation:

\begin{equation}
I_{theor}=n I_{theor} (1-e^{-\frac{\tau_{theor}}{RC}}),
\end{equation}

\noindent which leads to the following estimation of the moment when the current source produces the threshold value:

\begin{equation}
\tau_{theor}=-RC \cdot ln(\frac{n-1}{n})
\end{equation}

Thus, in a very simplified approximation, we come to the conclusion that in a heat-free model with a realistic pump pulse and neglected photon population dynamics inside the cavity, the build-up time assessed using only pump curvature properties is a constant equal to the product of the rise time constant of the current pulse and the specified emission level $\alpha$ when the laser is considered to be turned on:

\begin{equation}
t_{\alpha}= \tau_\alpha - \tau_{theor} =  \alpha RC
\label{syntdel}
\end{equation}

If the photonic mode population evolution is infinitely fast, the provided formula assesses the lowest value of build-up time possible for the non-heating, non-zero rise time case.

\subsection{Joule heating and threshold characteristics}

\begin{figure}[h]
\centering
\includegraphics[scale=0.32]{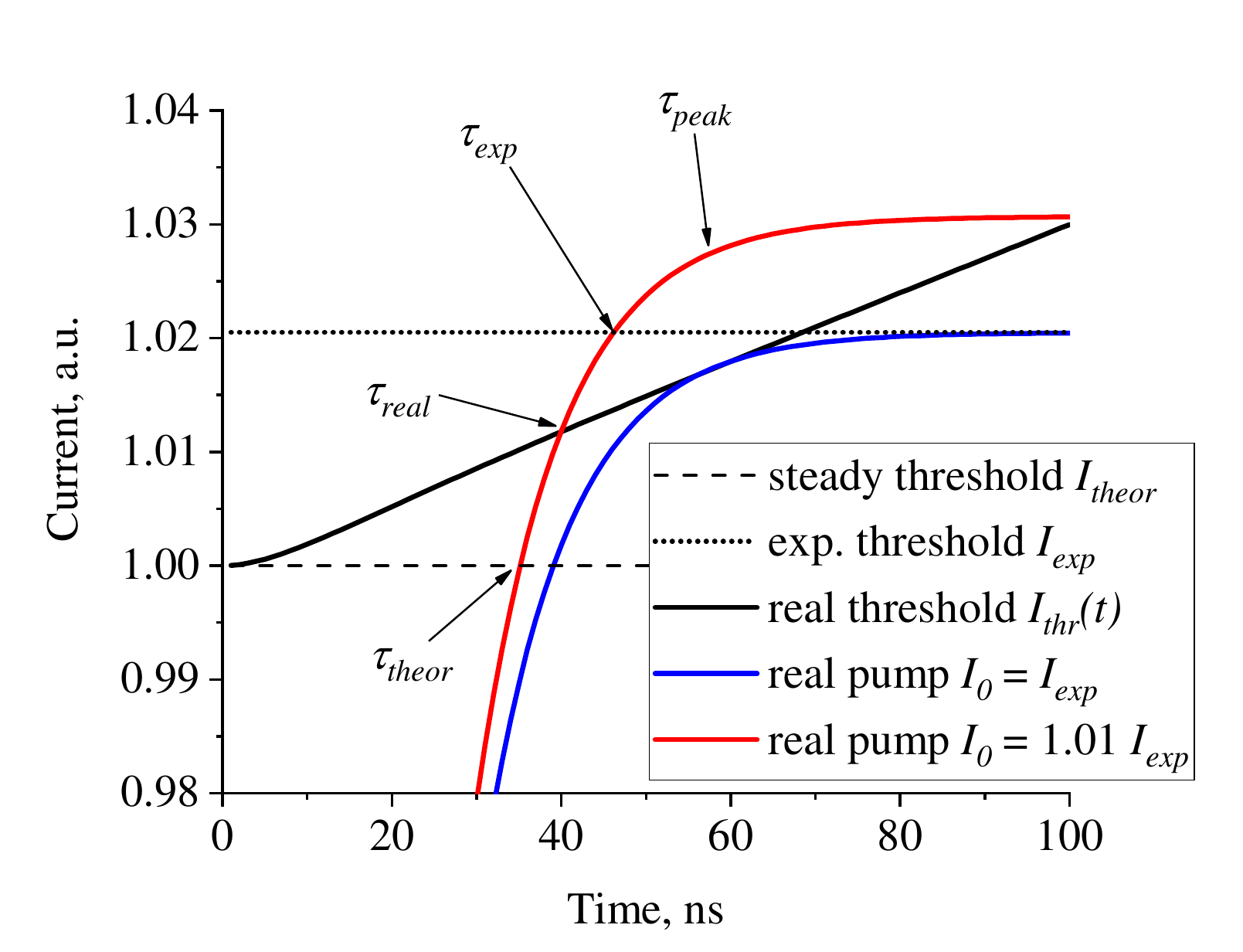}
\caption{Time profiles for pump pulses with close values of equilibrium current (blue and red solid lines) and time-dependent evolution of threshold current under Joule heating are given by eq.~\ref{eq:InstThres} (black solid line). Dotted and dashed lines designate two steady values of threshold: theoretical, valid for non-heating case, and the one that is expected to be experimentally observed. $\tau_i$ described in the main text denotes basic time moments used in the assessment of build-up time.}
\label{fig:mainsketch}
\end{figure}

Now we add heating processes to the model above.
In \cite{vrubel23} it was shown that the Joule heat inside the QCL active region does not dissipate from it at least for the first 50-100~ns. In adiabatic approximation, where the released energy is assumed to be enclosed inside the active region volume, that shifts the instantaneous threshold current according to the expression:

\begin{equation}
I_{thr}(t)=I_{theor}(1+\frac{U\int_{0}^t I_p(x)dx}{cm } \frac{1}{T_0})
\label{eq:InstThres}
\end{equation}

\noindent where $U$ [V] is an applied bias voltage, $cm$ [J~K$^{-1}$] is an active region heat capacity, $T_0$ is a constant determining the increase of threshold current with heating.

Fig.~\ref{fig:mainsketch} illustrates this shift: let us consider the experimental conditions. While the black solid curve indicates the threshold current given by the expression \ref{eq:InstThres}, the dotted horizontal line is the value, given by the power source and indicating the minimum equilibrium current value ($I_0$), which allows pump current to achieve time-dependent threshold current ($I_{thr}(t)$). So, it is clearly seen that in the experiment, at the very beginning of the pump pulse, the threshold current (black dashed line, $I_{theor}$) is lower than the one that can be formally obtained from the power source when measuring e.g. L-I-curve.
The solid red line in Fig.~\ref{fig:mainsketch} indicates the current, which is formally 1\% higher than the experimentally determined one ($I_{exp}$). Two points are of importance here: the first is the moment ($\tau_{real}$) where the 1.01 pump pulse initiates the emission. The second is a moment ($\tau_{exp}$) when the pump pulse formally achieves the threshold current given by the L-I-curve ($I_{exp}$).

The moment when a laser achieves maximum of its emission ($\tau_{peak}$) can be obtained by a solution of the following transcendental equation:

\begin{equation}
\frac{d(I_p(t)-I_{thr}(t))}{dt}=0
\end{equation}

In the case of adiabatic heating, the moment of $\alpha$-fraction of the maximum output power emission ($\tau_\alpha$) is calculated based on the intensity peak value in the following manner:

\begin{equation}
I_p(\tau_\alpha)-I_{thr}(\tau_\alpha) = \alpha (I_p(\tau_{peak})-I_{thr}(\tau_{peak})) 
\label{eq:heatingtransceq}
\end{equation}

\begin{figure}[h!]
\centering
\includegraphics[scale=0.32]{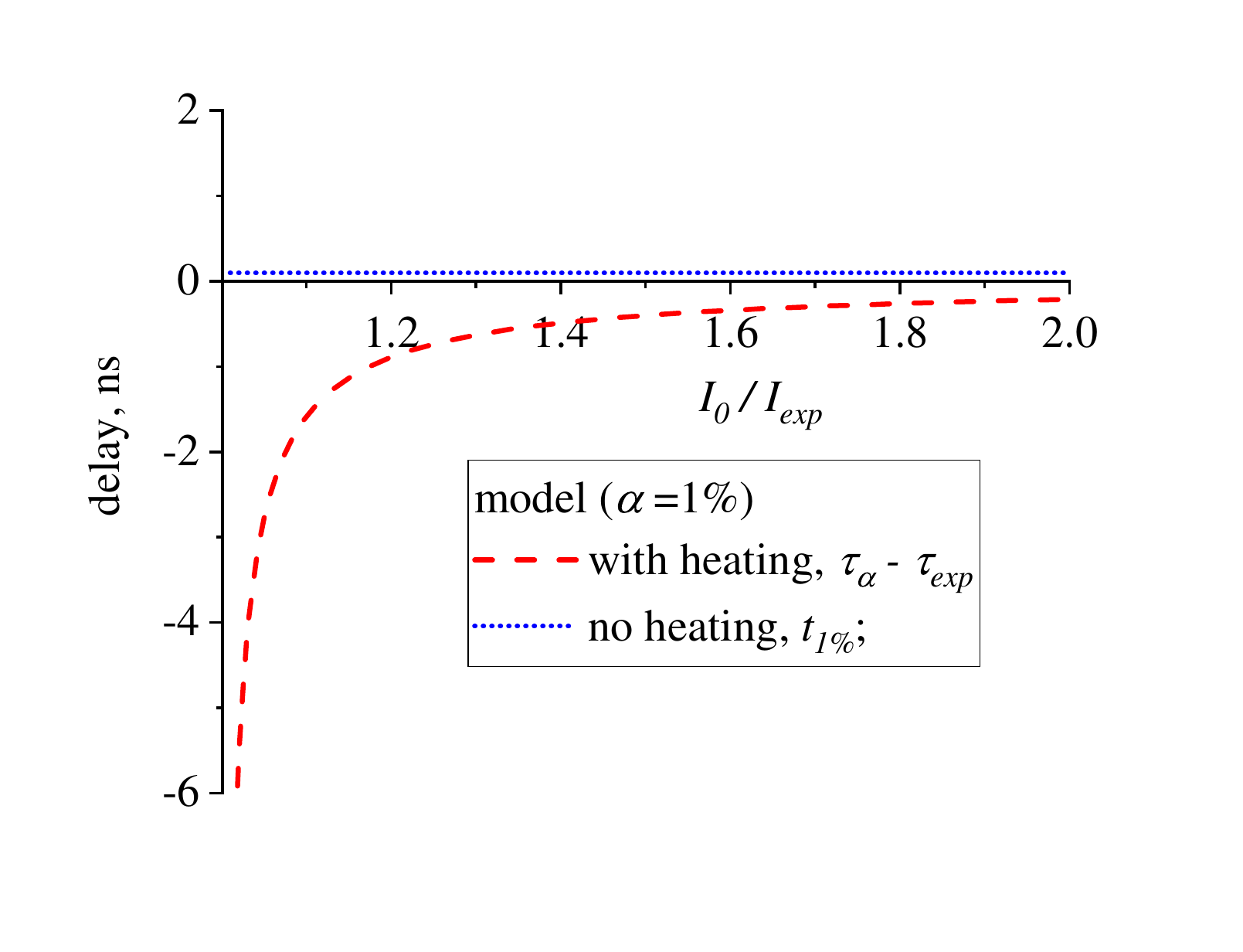}
\caption{Build-up time calculated by means of eq.~\ref{eq:heatingtransceq} with thermal effects included (red dashed line) and~\ref{syntdel}, indicating the minimum time necessary for power source to achieve pumping power sufficient for generation of $\alpha$-fraction of maximum output power (blue dotted line).
}
\label{fig:delay}
\end{figure}

With these equations, the build-up time can be numerically calculated and is depicted in Fig.~\ref{fig:delay}. The blue dotted line shows the non-heating model result from the previous section, while the red dashed line takes into account the thermal shift of the threshold current.

So, while in the non-heating case the build-up time is just a product of the power source time constant and the fraction of maximum emission, the temperature brought to the consideration reveals new features in the build-up time-current dependence. While the pumping is low enough, the threshold, formally taken from the L-I curve measurements, is actually higher than the real time-dependent threshold current, see $I_{thr}(t)$ in Fig.~\ref{fig:mainsketch}. So negative values of the delay time can easily appear. 
Moreover, one can see the asymptotic vanishing of the red curve at higher pump currents, which is explained by the decreasing effect of self-heating for the fast-rising edge.

\subsection{Photon mode population dynamics}

Although the previous considerations are based on quite reliable experimental assumptions, the latter result significantly differs from that of one can either find in experiment\cite{Cherotchenko22obs} or obtain by the rate-equation modeling\cite{Hamadou09}. The main difference lies in the range of pumping currents close to the threshold value, where the rate equation approach suggests the build-up time grows monotonously and the experiment reveals the non-monotonous behavior of the build-up dependence. To resolve this contradiction, we take into account the effect of photon mode population dynamics.

\subsubsection{Heating-less case, basic theory}

Photon population mode dynamics is governed by well-known equation\cite{Hamadou09}

\begin{equation}
\frac{dN}{dt} = \Gamma c'\sigma  (u-m) N + V \beta \frac{u}{\tau_{sp}} - \frac{N}{\tau_{loss}}
\label{eq:hamadoubasic}
\end{equation}

\noindent where $\Gamma c'\sigma$ is a set of design and material-related constants, $u$ and $m$ are concentrations of electrons of upper and middle level, $N$ is a number of photons in the cavity, $\beta$ is a spontaneous emission coefficient, $V$ is a volume of an active region, and $\tau_{loss}$ is a lifetime of photons in a cavity due to emission and absorption in the cavity media.

To evaluate photon population evolution in the case when $I_p>I_{thr}$, one needs to consider some specific parameters intrinsic to (i) spontaneous emission operation ($I_p<I_{thr}$) and (ii) the case when $I_p = I_{thr}$ e.g. transparency mode.

(i)The equilibrium under-threshold pumping current causes the emission of spontaneous radiation characterized by a steady number of photons in the cavity. Due to the stability of the photonic population ($\frac{dN}{dt} = 0$) and negligibly small stimulated processes, the latter reads as:

\begin{equation}
N= V \beta u \frac{\tau_{loss}}{\tau_{sp}}
\end{equation}

Substituting values from \cite{Hamadou09} one can assess the number of photon in spontaneous mode as:

\begin{equation}
\overline{N}_{sp}\approx 10^2
\end{equation}

which in the steady state is equivalent to emission with the power of 

\begin{equation}
L_{sp}\approx 1\text{$\mu$W}
\end{equation}
These values are only estimates and depend on the particular laser structure; anyway, this assessment allows one to understand the general behavior of the system. 

(ii)The pumping current having an infinitesimally small excess above the threshold results in the regime where losses of photons are equal to the stimulated emission. Considering the steady state, it is convenient to characterize the value of population inversion using eq.~(\ref{eq:hamadoubasic}) neglecting the spontaneous term:

\begin{equation}
(u-m)^\ast= \frac{1}{\Gamma c' \sigma \tau_{loss}}
\end{equation},
\textcolor{black}{where $(u-m)^\ast$ denotes the critical population inversion}.

While the characteristic time of the power source lies in the nanosecond domain, the primary effects of electronic relaxation in the cavity are the phonon-assisted processes, whose rates are characterized by picoseconds\cite{Hamadou09}. It means that processes related to the current variation can be considered quasi-stationary, which means that the electronic concentrations follow the current pulse profile. The latter allows to rewrite the population inversion for eq.~(\ref{eq:hamadoubasic}) in analytic approximation, neglecting spontaneous emission:

\begin{eqnarray}
\frac{dN}{dt} = \Gamma c'\sigma  (u(t)-m(t)) N - \frac{N}{\tau_{loss}} = \\ = \Gamma c'\sigma  (u-m)^\ast \frac{I_p(t)}{I_{thr}} N - \frac{N}{\tau_{loss}}
\label{eq:staticpopapprox}
\end{eqnarray}

This approximation is valid before photonic mode population rises and reduces population inversion to the critical value of $(u-m)^\ast$ in steady emission operation.
Finally eq.~(\ref{eq:staticpopapprox}) can be simplified as follows:

\begin{equation}
\frac{dN}{dt} =  \frac{1}{\tau_{loss}} \left(\frac{I_p(t)}{I_{thr}}  - 1 \right) N.
\label{eq:finalbtheor}
\end{equation}

\subsubsection{Heating less case, Heaviside pumping}

We start from the step-like pumping, which is discussed in \cite{Hamadou09}. The beginning of the pump pulse causes the electronic level population inversion to be higher than the critical value with the time typical for phonon-assisted processes of $\sim$picoseconds. Thus, the photonic mode evolution before achieving the $\alpha$ threshold can be rewritten as follows:

\begin{equation}
\frac{dN}{N} =  \frac{1}{\tau_{loss}} \left(\frac{I_0}{I_{theor}}  - 1 \right) dt
\label{eq:expdiffeq}
\end{equation}
 
After simplification, one can find the time interval needed to gain the number of photons corresponding to the $\alpha$ threshold of output power for the given pump current:

\begin{equation}
t_{\alpha} =  \frac{\tau_{loss}}{ \frac{I_0}{I_{theor}}-1 } ln\left( \frac{\alpha \times \overline{N}_{I_0}}{\overline{N}_{sp}}\right).
\label{eq:HSannod}
\end{equation}

The number of photons occupying a laser cavity emitting steady radiation can be assessed via the light-current characteristic in the following manner:

\begin{equation}
\overline{N}_{I_0} = \frac{\chi*I_{theor}(\frac{I_0}{I_{theor}}-1) }{e^-  \hbar\omega} \tau_{out}
\label{eq:eqphotmodepop}
\end{equation}

\noindent where $\chi$ is a light-current characteristic slope, $e^-$ is the electron charge, $\hbar\omega$ is a photon energy, $\tau_{out}$ is a mirror loss. With this estimation one can get the final equation:

\begin{equation}
t_{\alpha} =  \frac{\tau_{loss}}{ \frac{I_0}{I_{theor}}-1 } ln\left( \frac{\alpha \times \chi I_{theor}(\frac{I_0}{I_{theor}}-1) \tau_{out} }{e^-  \hbar\omega \overline{N}_{sp}} \right).
\end{equation}

Adopting typical values of laser parameters as in \textcolor{black}{Table~\ref{tbl:summary}} one can assess the population of photonic mode at the level of 1\% threshold as:

\begin{equation}
\overline{N}_{I_0} \approx 10^8 (\frac{I_0}{I_{theor}}-1)
\end{equation}

The estimated occupation numbers of a cavity mode differ drastically for spontaneous and working regimes; however, the exponential evolution of photonic mode occupation, \textcolor{black}{ according to eq.~\ref{eq:expdiffeq}, diminishes these differences.} In fact, the natural logarithm vanishes these orders of magnitude difference in the following manner:

\begin{equation}
ln(\frac{10^{-2} \overline{N}_{1.01I_{theor}}}{\overline{N}_{sp}})\approx 5 \text{ and } ln(\frac{ 10^{-2}\overline{N}_{2.00I_{theor}}}{\overline{N}_{sp}})\approx 10.
\end{equation}

Blue line in Fig.~\ref{fig:HSandRC} indicates this result, where $t_\alpha$ is mainly inversely proportional to $\frac{I_0}{I_{theor}}-1$. 

\subsubsection{Heating less case, smooth pumping edge}

To move on, we consider the non-zero rise time of pumping pulse in the non-heating case, when the temperature of the active region remains constant.
In this case, the population inversion in eq.~(\ref{eq:finalbtheor}) can be approximated by a linear function increase starting from the critical value $(u-m)^\ast$ with the rate defined by the time constant of a power source: 

\begin{equation}
\frac{dN}{N} =  \frac{1}{\tau_{loss}} \left(\frac{I_{theor}+I_p'(\tau_{theor})\times t}{I_{theor}}  - 1 \right) dt
\label{eq:diffeqforfinrise}
\end{equation}

\noindent The solution obviously reads:

\begin{equation}
N=\overline{N}_{sp} exp^{\frac{I_p'(\tau_{theor})}{I_{theor}} \frac{t^2}{2\tau_{loss}} }
\end{equation}

\noindent After simplification one can easily get:

\begin{equation}
 t_{\alpha} =  \sqrt{ 2 \tau_{loss} \frac{I_{theor}}{I_p'(\tau_{theor})} ln\left(\frac{\alpha \times \overline{N}_{I_0}}{\overline{N}_{sp}}\right)   }
 \label{eq:RISEanmodel}
\end{equation}

Considering the realistic pump current (see eq.~\ref{eq:realisticpump}), we can write explicitly the derivative of a pump pulse at the moment $\tau_{theor}$ with the final result for build-up time:

\begin{equation}
t_{\alpha} =  \sqrt{ \frac{2 \tau_{loss} RC } {\frac{I_0}{I_{theor}}-1} ln\left(\frac{\alpha \times \overline{N}_{I_0}}{\overline{N}_{sp}}\right)  }
\label{eq:EXPanmodel}
\end{equation}

\noindent where $\overline{N}_{I_0}$ is again photon population in the steady state, as assessed in eq.~(\ref{eq:eqphotmodepop}).

\begin{figure}[h!]
\centering
\includegraphics[scale=0.3]{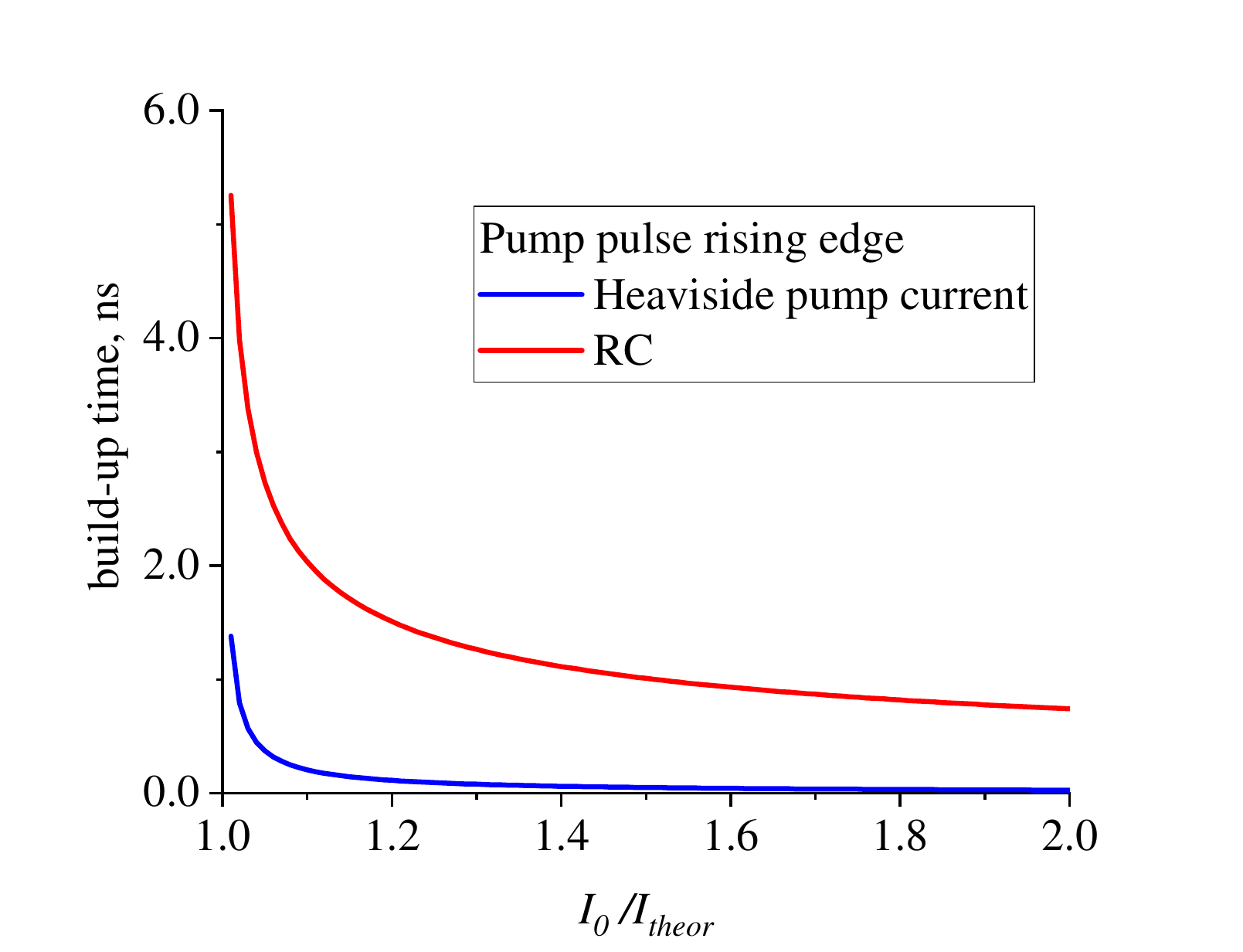}
\caption{Analytical estimations of a build-up time dependence on pump current. Red line depicts the curve, calculated \textcolor{black}{using eq.~\ref{eq:EXPanmodel}} for pumping pulse with finite rise time characterized by typical time constant RC, \textcolor{black}{see eq.~\ref{eq:realisticpump}, while blue line } indicates the build-up time characteristic for Heaviside-like current pulse\textcolor{black}{, see eq.~\ref{eq:HSannod}}. Basic parameter $\tau_{loss}$ of 3~ps common for both formulas is adopted from Ref.~\cite{Hamadou09} }
\label{fig:HSandRC}
\end{figure}

Red curve in Fig.~\ref{fig:HSandRC} illustrates this dependence for the non-heating case. 

\section{discussion and conclusions}

\subsection{Fundamental peculiarities of build-up time measurements}

The use of methodology to measure and interpret the build-up time for QCL in the experiment raises some fundamental questions, which are of importance especially for high-threshold devices.
The basis of these questions is a ``self-interaction''-like effect: on the one hand, the value of a build-up time is a function of pumping current; on the other hand, varying pumping current creates variable conditions, which in turn affect the build-up time in a self-consistent manner.
Build-up time plot measured assuming that the threshold current is an experimentally observed constant (see $I_{exp}$ in Fig.~\ref{fig:mainsketch}) may exhibit an artifact such as the negative delay time shown in Fig.~\ref{fig:mainsketch}.

Analysis of the contribution of the thermal processes to the build-up time in the form of synthetic negative time shift is obscured due to its transcendental nature. One can easily see that it monotonously decreases in the near-threshold limit.

Analytic conclusions related to photonic mode occupation evolution are more fruitful.
Synthetic example where the pumping front edge is characterized by the Heaviside function depends mainly on photon lifetime and pulse amplitude. In this model, the build-up time is simply proportional to the photon lifetime and, on the general scale, inversely proportional to the excess of pumping current above the threshold.

Realistic pulse with the exponential rising edge complicates the situation a bit. In this model, the time constant of the power source directly forms the build-up time value as a co-factor of photon lifetime. The square root of this product forms a dimensional part of the build-up time. Finally, this value is normalized by the square root of the excess pumping current above the threshold. Such a significant difference between participation of the excess of pumping current above threshold in the form of $\frac{I_0}{I_{thres}}-1$ in formulas for synthetic and realistic cases is dictated by the very different ways (linear or parabolic time dependence in the exponent) to achieve and exceed the critical population inversion value..

\subsection{Further development: from ultimate analytic model to numerical simulations}

\subsubsection{Near threshold correction}

\begin{figure}[h!]
\centering
\includegraphics[scale=0.3]{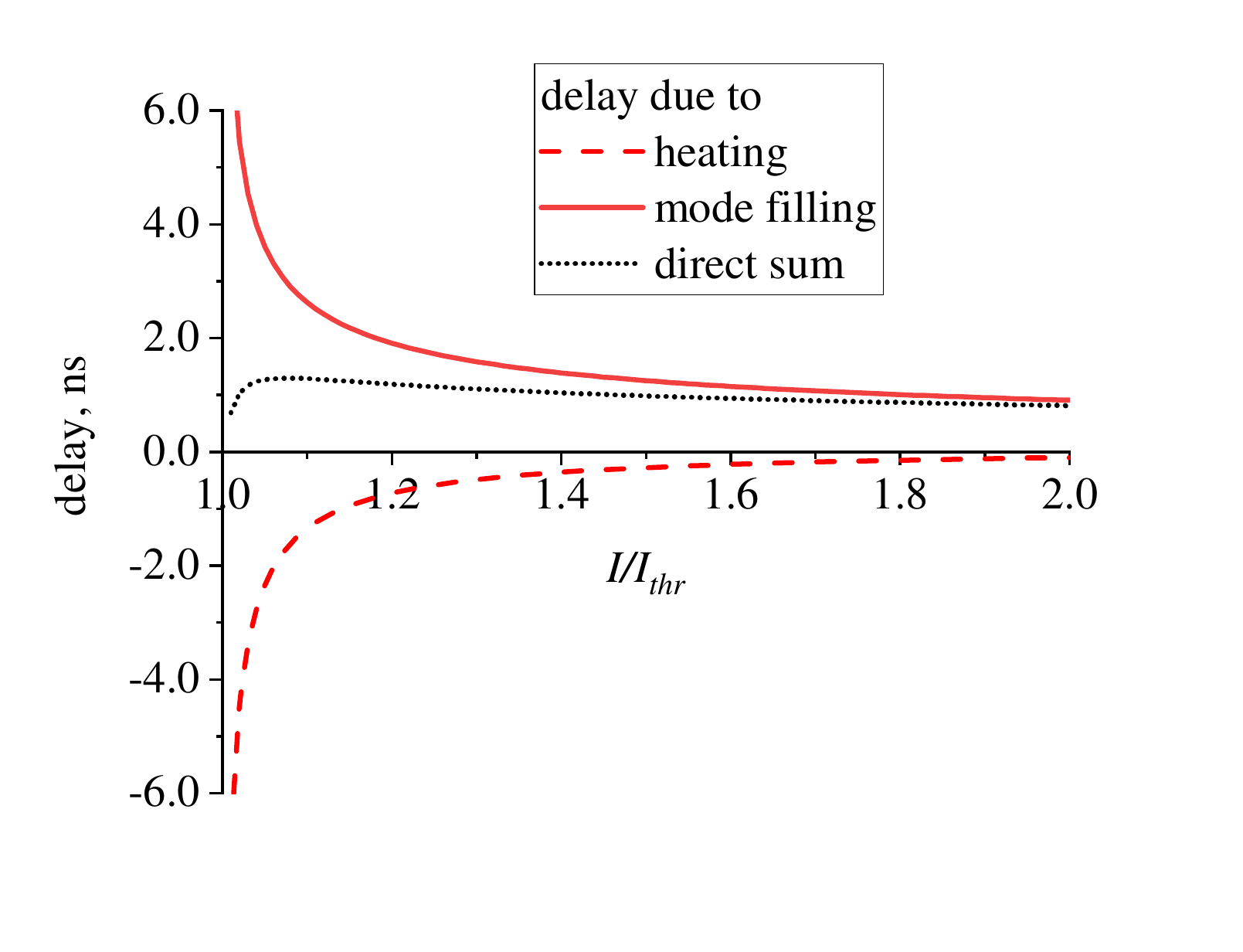}
\caption{Red solid line depicts the time needed for photonic mode to gain its population up to the limit when a laser emission achieves the $\alpha$-fraction of the maximum intensity as shown in Fig.~\ref{fig:HSandRC}. 
Red dashed line is taken from Fig.~\ref{fig:delay} and shows an effect of active region heating manifesting as a negative time shift when pump pulse has a non-zero rise time. 
The direct sum of these two processes depicted as black dotted line symbolizes possible experimentally observed build up time}
\label{fig:DELtotal}
\end{figure}

To introduce the ultimate analytic expression one should merge all the models accounting  for the heating of an active region with the model accounting for non-zero pumping rise time. In a first approximation it is possible to assess build-up by calculating direct sum using relevant curves depicted in  Fig.~\ref{fig:delay}~and~\ref{fig:HSandRC}. The result of such math is presented in Fig.~\ref{fig:DELtotal}, see blue line.

The applicability of such an approach is valid on the general scale, however, a further refinement can be proposed. If the threshold current increases, the eq.~\ref{eq:RISEanmodel} can be upgraded in a rather simple manner. 
The rate of photonic mode population increase \textcolor{black}{in eq.~\ref{eq:diffeqforfinrise} is derived under} assumption that the rate of electronic population inversion is proportional to the derivative of the pump current $I_p'(\tau_{theor})$. Obviously, this approach is fully correct in the heating-less case, when the threshold current is constant. When heating is considered, the data depicted in Fig.~\ref{fig:mainsketch} strongly suggests that the rate of electronic population inversion will be attenuated by growing threshold current. Approximate modification of the build up time assessment can be considered as follows:

\begin{equation}
t_{build-up} =  \sqrt{ 2 \tau_{loss} \frac{I_{thr}(\tau_{real})}{I_p'(\tau_{real})-I_{thr}'(\tau_{real})} ln\left(\frac{N}{\overline{N}_{sp}} \right)   }
\label{eq:nonzerothrslope} 
\end{equation}

Obviously, the closer pump current to threshold value is the smaller a slope of an electronic population inversion that feeds a photonic mode. Consequently, even a tiny slope of threshold current may play a role and distort build-up time plot in $I_0\rightarrow I_{thresh}$ limit. At high pump currents the effect of non-zero derivative of threshold current creates tiny perturbation of the build-up time compared to the one calculated via direct sum of delay due to heating and mode filling (blue line in Fig.~\ref{fig:DELtotal}). 

\subsubsection{Applicability of adiabatic approximation}

In this work effect of thermal processes on the build-up time is considered in adiabatic manner. Heat produced during pump pulse is considered enclosed in the volume of active region according to eq.~\ref{eq:InstThres} which is confirmed in \cite{vrubel23}. This approach is valid in the very beginning of pump pulse, until thermal distribution around active region becomes able to evacuate heat effectively. Generally, precise calculation of thermal distribution requires solution of time-dependent heat equation with obscuring effective parameters of media and non-trivial geometric configuration. Taking into account the fact that thermal conductivity of an active region is significantly lower than that of claddings, a significant temperature gradient inside cascades is expected\cite{vrubel23}. This means that under pumping each cascade is in individual conditions, that makes results of any homogeneous model approximate.


Finally, a complex interplay between thermal, 
electronic and optical processes in an active region under pumping and high spatial inhomogeneity of typical quantum cascade laser design suggest that fine estimation of build-up time for the given sample requires accurate numerical modelling.

\clearpage
\newpage
\section*{funding}
I.V. and E.C. acknowledge the support from the Russian Science Foundation (project 23-29-00930).

\section*{Author contributions}
All authors have accepted responsibility for the entire content of this manuscript and consented to its submission to the journal, reviewed all the results and approved the final version of the manuscript. IV and EC developed the theory and prepared the manuscript with contributions from all co-authors, and SA and KK developed the model code and performed the simulations. VD and GS supervised the project and edited the manuscript.

\section*{Conflict of interest}
Authors state no conflict of interest.

\section*{Data availability statement}

Data sharing is not applicable to this article as no datasets were generated or analyzed during the current study.

\newpage
\appendix

\section{Parameters used for the numerical assessments}
\begin{table}[h!]
\caption{The list of parameter values typical for QCL used throughout the work}
\begin{tabular}{c|c|c}
parameter & description & value \\ \hline
$\tau_{loss}$ & lifetime of photon in the cavity & 3~ps  \\
RC & power supply time constant  & 10~ns \\
$\chi$ & light-current characteristic slope & 0.5~W/A \\
$I_{theor}$ & threshold current & 0.5~A \\
$\hbar\omega$ & photon energy & 0.15~eV \\
$U$   &  bias voltage  &  10~V  \\
$cm$   &  active region heat capacity  & 4$\cdot$10$^{-7}$ J/K  \\
$T_0$  &  threshold current temperature coeff  &  200~K \\
$V$   & volume of active region   & 2.4$\cdot$10$^{-7}$~cm$^{3}$  \\
$\beta$   & \textcolor{black}{spontaneous emission factor}  & 0.002  \\
\end{tabular}
\label{tbl:summary}
\end{table}

\newpage

\bibliography{bibliography}

\end{document}